\documentclass[%
reprint,showkeys,
nofootinbib,  
amsmath,amssymb,
aps]{revtex4-2}

\usepackage{graphicx}
\usepackage{dcolumn}
\usepackage{bm}
\usepackage{siunitx}
\usepackage{xcolor}
\usepackage{balance}
\usepackage{siunitx}
\usepackage{bbding}
\usepackage[version=3]{mhchem}
\usepackage{mathptmx,amsmath,amssymb}

\begin{document}
	\title{Molecular adsorption induces normal stresses at frictional interfaces of hydrogels\\}



	\begin{abstract}
		{Lola Ciapa, Yvette Tran, Christian Fr\'etigny, Antoine Chateauminois, Emilie Verneuil \\
			{\it \noindent Soft Matter Science and Engineering (SIMM), CNRS UMR 7615, ESPCI Paris, PSL University, Sorbonne Universit\'e, F-75005 Paris, France\\}}

	Friction experiments were conducted on hydrogel thin films sliding against a rigid sphere in a low velocity regime where molecular adsorption at the sliding interface sets the friction force, through a dissipative adsorption-stretching-desorption mechanism initially postulated by Schallamach. By carefully imaging the contact from the initial indentation step of the sphere into the hydrogel to steady state sliding, we evidence for the first time that this very same adsorption mechanism also results in a normal force pulling the sphere further into the hydrogel. Observations of this tangential-normal coupling is made on a variety of chemically modified silica spheres, over 3 decades in velocity and at varied normal load, thereby demonstrating its robustness. Quantitative measurements of the extra normal force and of the friction-velocity relationship versus normal load are well rationalized within a theoretical model based on the thermal actuation of molecular bonds. To do so, we account for the finite non-zero thickness of the sliding interface at which molecular adsorption and stretching events produce an out-of-plane force responsible for both friction and normal pull-in.

\end{abstract}

	\maketitle


\section{Introduction}
Sliding contacts involving soft matter systems such as rubbers or gels encompass a wide variety of engineering and natural settings such as grippers\cite{wu2022}, soft robots~\cite{chen2020,whitesides2018}, tires on roads\cite{heinrich2008,lorenz2013}, contact lenses~\cite{Dunn2013,Rennie2005,Roba2011} or substitutes for cartilages~\cite{blum_low_2013,han_effect_2018,porte_lubrication_2020}. Although a control of sliding friction in these applications is highly desired, the complex interplay between a wide variety of dissipative mechanisms occurring over a wide range of length-scale makes the manipulation of friction a challenge.\\	
At a molecular scale, frictional dissipation in dry, unlubricated, sliding contacts with rubbers or gels is often assumed to be mediated by the pinning, stretching and subsequent de-pinning of polymer chains on the mating surface according to a thermally activated rate process. This seminal idea was first introduced by Schallamach~\cite{schallamach_theory_1963} to account of the bell-shaped velocity-dependence of rubber friction on smooth, hard surfaces which was reported in an early experimental work by Grosh.~\cite{Grosch1963} It was further revisited and extended in many theoretical works to incorporate, as an example, the effects of chain stiffness and viscoelasticity,~\cite{singh_steady_2011} the role of the aging of bonds~\cite{singh_model_2021} or normal load contributions~\cite{chernyak_theory_1986} in both steady~\cite{Shukla2024} and unsteady sliding situations~\cite{singh_model_2021,dawara2024}. Interestingly, this description of sliding friction also shares some common features with dynamic spectroscopy experiments~\cite{friddle_interpreting_2012,friddle_near-equilibrium_2008} or with the motions of cells~\cite{Woessner2024,Sens2020}. Beyond rubber friction~\cite{vorvolakos_effects_2003,singh_model_2021}, It was also extended to the friction of self-assembled monolayers~\cite{drummond_friction_2003} and of polymer brushes\cite{ohsedo_surface_2004}.\\   
In the case of water swollen hydrogel networks, frictional energy dissipation may be complicated by the interplay between the above mentioned interfacial pinning/depining mechanisms, elasto-hydrodynamic effects and poroelastic dissipation.~\cite{shoaib_advances_2020,gong2006,bonyadi2020} In an extended set of experiments, Gong and co-workers observed a non monotonic trend of the friction force which is first increasing and then decreasing with sliding velocity. After the achievement of a minimum in the friction force, friction is then re-increasing with velocity, which was attributed to a transition to elasto-hydrodynamic lubrication (EHL).~\cite{gong_gel_1998,Kurokawa2005,Kagata2002} In the case of gel systems having attractive interactions with the mating surface, this behavior is described within the framework of modified form of the Schallamach's model.~\cite{gong_gel_1998,Kurokawa2005} Here, the frictional stress is assumed to include two additive contributions, namely (\textit{i}) the elastic stretching of adsorbed polymer chains according to the original Schallamach's model and (\textit{ii}) the shear of the viscous solvent within a superficial gel layer at the vicinity of the interface. According to this model, below a characteristic velocity $v_f$ depending on the viscosity of the solvent and on characteristic times for polymer attachment and detachment, the first contribution is dominant and the frictional stress is predicted to increase logarithmically with the sliding velocity. Above $v_f$, the second contribution becomes predominant and a transition to hydrodynamic lubrication is predicted. This model provides scaling laws that qualitatively account for many experimental results where, as an example, the hydrophobicity of the substrate~\cite{tominaga_effect_2008}, the composition of the hydrogel network~\cite{shoaib_insight_2018}, its crosslinking density~\cite{gong2006,gong_synthesis_2001,li2012} or the nature of the solvent are changed.~\cite{cuccia_pore-size_2020,nakano_effect_2011} {However, the large number of adjustable parameters prevents this analysis from being quantitative.}\\
In order to quantitatively characterize the dissipative effects of molecular adsorption at the sliding interface, we offer to work in the low velocity regime $v\ll v_f$ where any viscous contribution to dissipation can be neglected.
In a previous study, we  investigated the frictional response of thin poly(dimethylacrylamide) gels films in contact with spherical glass probes using a dedicated torsional setup allowing to achieve an homogeneous sliding velocity within the contact interface while discarding the effects of poroelastic flow and elasto-hydrodynamic lubrication.~\cite{ciapa_friction_2024} In a regime where the frictional stress was observed to vary logarithmically with the sliding velocity, we showed that the Schallamach's model allows to describe quantitatively the load- and velocity-dependence of the frictional stress. Additionally, the model provides adhesion energies and characteristic times for molecular adsorption that are found consistent with the physical chemistry of the chemically-modified silica surfaces.\\
In the present study, we report on new experimental observations using the same system where we evidence for the first time a coupling between normal and lateral forces which results in a \textit{re-embedding} of the glass probe into the gel when sliding is initiated {together with a non-trivial dependence of friction with contact area}. This coupling is discussed within the framework of a modified Schallamach's model derived from the theory of Chernyak and and Leonov.~\cite{chernyak_theory_1986,leonov_dependence_1990} This approach takes into account the non vanishing thickness of the layer of adsorbed surface polymer chains, where molecular adsorption and stretching events produce an out-of-plane force responsible for both friction and normal pull-in.\\
In a first section, we provide a theoretical background of the Schallamach's model. Then, experimental details and results obtained over a wide range of sliding velocities and normal forces are presented for spherical probes with varied physical chemistry. In a last section, we derive our friction model for finite interface thicknesses and we show that it  accounts for the experimental observations in a semi-quantitative way.\\

\section{Theoretical background on thermally-activated frictional dissipation}

\begin{figure}[!ht]
	\includegraphics[width=8cm]{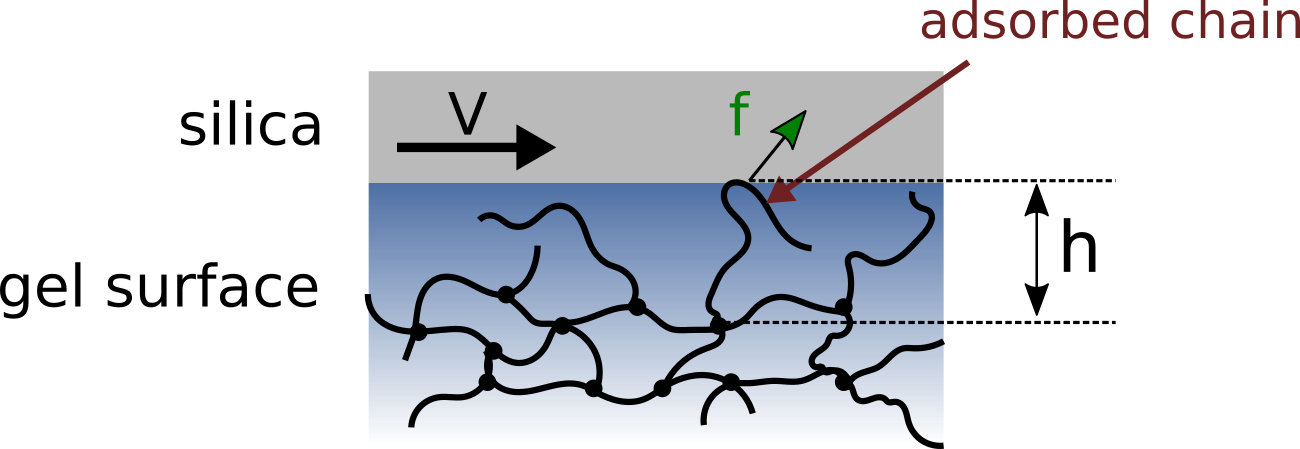}
	\caption{Schematics of an interface between a polymeric gel and a solid sliding at velocity $V$. Polymer chains adsorption is timed by a thermally-activated mechanism. Once adsorbed, a stretched chain experiences a force $f$ with both a lateral and normal component. The distance between the solid and the bulk gel is denoted $h$. As $h\rightarrow 0$, $f$ is directed in the interface plane and its normal component vanishes. Else, a non-zero $f$-normal component acts on the interface.  }
	\label{fig:Schallamach_leonov}
\end{figure}

The frictional properties of hydrogel thin films of polydimethylacrylamide sliding against chemically functionalized silica spheres were recently showed to be set, in a peculiar velocity regime, by the molecular adsorption of the hydrogel polymer chains through a dissipative adsorption-stretching-desorption mechanism~\cite{ciapa_friction_2024} schematized in Fig.~\ref{fig:Schallamach_leonov}. As mentioned in the introduction, this thermally activated dynamics was first introduced by Schallamach for rubber friction.~\cite{schallamach_theory_1963}.
We first recall the main features of the formalism developed to model the thermally-activated regime for the friction at the hydrogel/silica interface, where the interface is first considered of vanishing thickness $h \rightarrow 0$ (Fig.~\ref{fig:Schallamach_leonov}). \\
The polymer chains at the interface are likely to adsorb at the silica interface. Adsorption is then characterized by the mean lifetime of a bond at rest $\tau_{\mathrm{ads}}$, when sliding velocity is $V=0$, and the lifetime of a free chain $\tau$ where $\tau<<\tau_{\mathrm{ads}}$ since adsorption is thermodynamically favored. The ratio between these two mean residence times is set by the energy barrier $W$ separating the bond and free state through an Arrhenius law $\tau_{\mathrm{ads}}/\tau=e^{W/k_BT}$. The survival probability of a bond after a time $t_a$ is denoted $P_{0,0}(t_a)$ where the subscripts denote the two conditions: zero sliding velocity and zero interface thickness: $P_{0,0}(t_a)=e^{-{\frac{t_a}{\tau_{\mathrm{ads}}}}}$.\\
During sliding at velocity $V$, the bonds are elastically stressed, resulting in a reduction of the energy barrier $W$ to depin. The model describes the lifetime of the bonds and the energy dissipated as they break by accounting for the bias in the energy landscape due to the elastic energy stored in the stretched chains when the bonds move at velocity $V$ along the interface plane.  Following Schallamach, the energy bias is assumed to be proportional to the elastic force $f$ on the bond through the activation length $\lambda$, set here by the hydrogel network mesh size. 
Hence, the polymer chains are also characterized by their stiffness $M$ which linearly relates their stretch to the force $f$. If a bond is translated along the plane of the interface at velocity $V$ for a time $t$, then $f(t)=MVt$. Here, we conveniently define $\ell$ as the characteristic stretching length when the elastic energy of the stretched chain $M\ell\lambda$ is equal to the thermal energy $k_BT$, where $k_B$ is the Boltzmann constant and $T$ the temperature : $\ell=\frac{k_BT}{\lambda M}$. Hence, the survival probability $P_{V,0}(t_a)$ for a bond formed at $t=0$ and existing at a time $t_a$ is:
\begin{eqnarray}
	P_{V,0}(t_a)=\exp\left( -\frac{1}{\tau_{\mathrm{ads}}}\int_0^{t_a}e^{\frac{\lambda f(t)}{k_BT}}\,dt \right)\\
	=\exp\left( -\frac{1}{\tau_{\mathrm{ads}}}\int_0^{t_a}e^{\frac{Vt}{\ell}}\,dt \right)
\end{eqnarray}
where the subscripts $V,0$ denote the non-zero sliding velocity and the zero interface thickness $h=0$.
From this, past works\cite{drummond_friction_2003, singh_steady_2011, ciapa_friction_2024} identified a  thermally activated regime for friction in the velocity range between $\ell/\tau_{\mathrm{ads}}$ and $\ell/\tau$, in which the average number of bonds is nearly constant with $V$ and equal to the number density of possible adsorption sites $N_0$, the bond lifetime is $\tau_{\mathrm{ads}}$, while the energy stored in the stretched chains increases with $V$, so that dissipation increases with velocity. The result is a logarithmically increasing frictional stress with sliding velocity $V$ which writes :
\begin{equation}
	\sigma_t=N_0 \frac{k_BT}{2\lambda}\ln\left(\frac{V}{\ell/\tau_{\mathrm{ads}}}\right)
	\label{eq:sigmat:model:singh}
\end{equation}
In this primary version of the model, note that the force $f$ is oriented within the plane of the interface whose thickness $h$ is vanishing. By varying the chemical interaction at stake at the hydrogel/glass interface, a previous study allowed to show that indeed, the friction stress varies as the logarithm of the velocity in a particular velocity range.~\cite{ciapa_friction_2024} It further allowed to measure the adsorption time $\tau_{\mathrm{ads}}$ consistently with the surface chemistry of the glass counterpart that sets the energy barrier of the adsorption mechanism. In the present paper, we further explore this thermally-activated regime by focusing on the interplay between frictional (tangential) stresses and normal stresses. We first report experimental observations evidencing several couplings between both components of the stress which point to the need to refine the current models. From these, a revised version of the adsorption-stretching-desorption model is offered which accounts for the non-zero thickness of the sliding interface. A comparison of the experimental data with the model is then discussed.  \\
%
\section{Materials and methods}
Friction experiments were carried out using poly(dimethylacrylamide) (PDMA) hydrogel films covalently grafted onto glass substrates. These films were synthesized by simultaneously crosslinking and grafting preformed ene-functionalised polymer chains onto glass substrates using a thiol-ene click reaction which is fully described elsewhere \cite{delavoipiere_poroelastic_2016, delavoipiere_friction_2018}. Strong adhesion between the hydrogel film and its substrate was achieved by functionalising borosilicate glass slides with thiol groups. Through a thiol-ene reaction, the gel film was covalently bound to the glass slide, thereby preventing interfacial debonding during swelling and friction. Two different PDMA films with slightly different thicknesses and swelling ratio were synthesized and their dry and swollen thicknesses were measured by spectroscopic ellipsometry. The swollen thickness $e_0$ and swelling ratio $S_{\mathrm{w}}$ are $e_0=2.8$~\si{\micro\meter} and $S_{\mathrm{w}}=2.3$ for one film, and $e_0=2.3$~\si{\micro\meter} and $S_{\mathrm{w}}=1.8$ for the other one.~\footnote{These low values of swelling ratio ensure that no wrinkles or creases form at the gel interface. Indeed, buckling instabilities have been described in the past as a result of compressive swelling stresses building up in surface-attached films, and were found \cite{trujillo_creasing_2008, hong_formation_2009, Chen_Crosby_Hayward_2014} to occur above a swelling ratio threshold $S^c_{\mathrm{w}}=2.40$ for creases and 3.40 for wrinkles, so that the swelling ratios of our two films stand below these thresholds.}
From the values of $S_{\mathrm{w}}$, we further estimate the number of Kuhn segments between crosslinks\cite{ciapa_friction_2024} $\nu_c=13$ and $\nu_c=6$ respectively, so that $\nu_c$ will be approximated as $\nu_c\sim 10$ in what follows. The stiffness $M$ of a single chain in the network is estimated from $M=k_BT/\nu_c b^2$ with $b=1$~nm the Kuhn length for PDMA. In the following, the equilibrium length $h_0$ of a chain at rest at the gel interface will first be considered equal to that between crosslinks. Hence, $\ell=\frac{k_BT}{\lambda M}=\frac{\nu_cb^2}{\lambda}$ can be estimated as $\ell\sim10$~nm.\\

Friction experiments we performed using fused silica lenses (Newport, UV fused silica SPX114) with curvature radius $R=23$~mm as probes. In order to vary their molecular interactions with PDMA hydrogel, the silica surfaces were grafted with three different silanes, namely aminopropyltriethoxysilane (APTES, ABCR GmbH), propyltriethoxysilane (PTES, ABCR GmbH) and octadecyltrichlorosilane (OTS, ABCR GmbH) using a protocol which is fully detailed elsewhere.~\cite{ciapa_friction_2024} In what follows, the PTES, APTES and OTS treated surfaces will be denoted as "propyl", "aminopropyl" and "octadecyl", respectively.\\

\begin{figure}[!ht]
	\includegraphics[width=8cm]{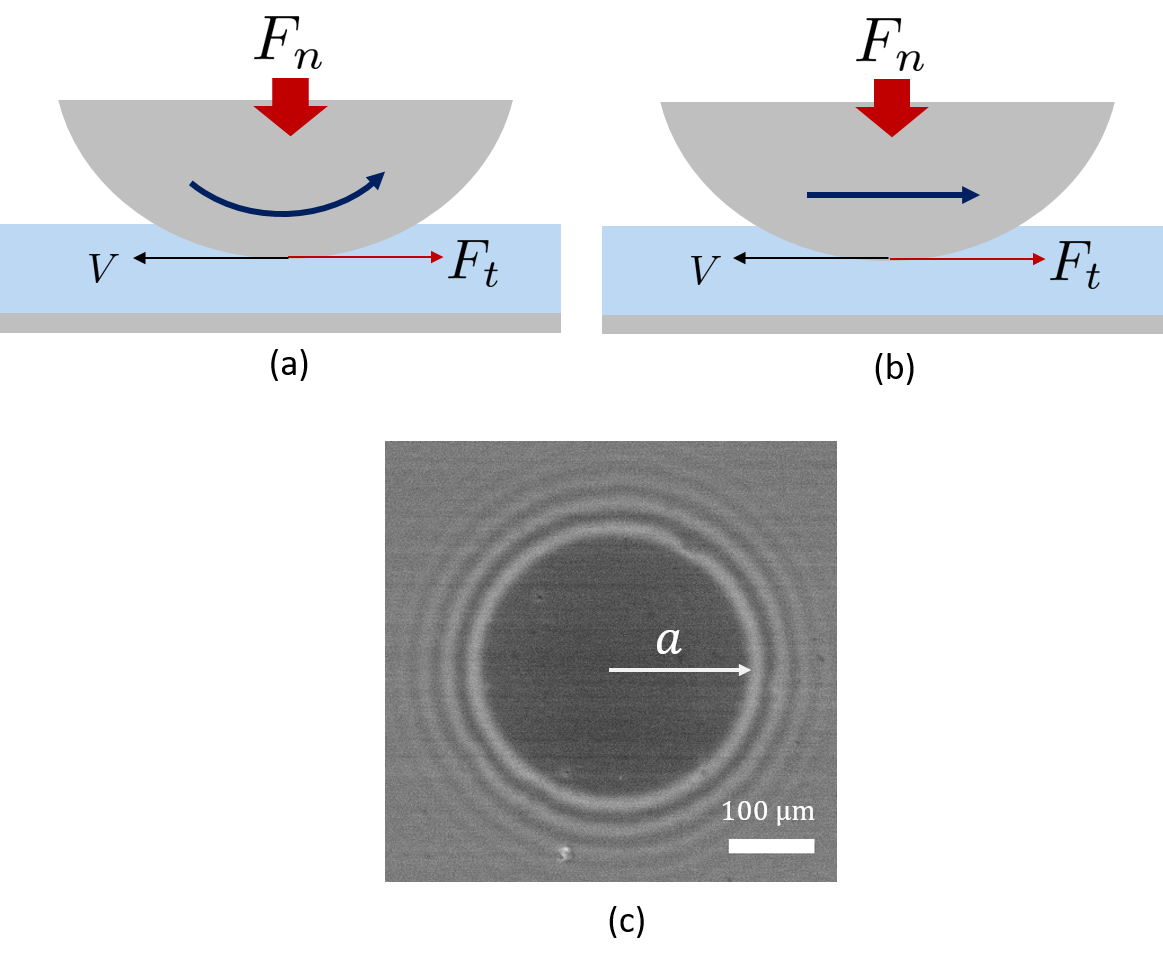}
	\caption{Schematic of the experimental set-ups: A lens of radius $R$ slides against a swollen hydrogel film with thickness $e_0$, immersed in water. Dry thickness is $e$. (a) The lens rotates (A) or is linearly translated with respect to the film. The interfacial velocity $V$ is uniform within the contact. A constant normal load $F_n$ is imposed. The friction force $F_t$ is measured from the deflection of cantilevers with calibrated stiffnesses. (c) Contrasted RICM images of the contact allow for accurate measurement of the contact shape and radius $a$ from which the indentation depth $\delta=a^2/(2R)$ is computed.}
	\label{fig:setup}
\end{figure}

Two different home-made setups differing in the contact kinematics conditions were used for the friction experiments. The first one is a rotational friction device fully described in reference~\cite{ciapa_friction_2024} where the lens is rotated (Fig.~\ref{fig:setup}a) in order to achieve linear sliding conditions within the contact interface which remains fixed with respect to the film. In the second device (Fig.~\ref{fig:setup}b), the film is translated with respect to the fixed lens along a rectilinear trajectory~\cite{delavoipiere_friction_2018}.
%
In steady-state, the main difference between these two situations is the poroelastic transport of water within the gel film forced by the sliding of the probe when it is translated (Fig.~\ref{fig:setup}b) and that does not exist in rotation (Fig.~\ref{fig:setup}a). Here, based on previous results~\cite{delavoipiere_friction_2018}, care was taken to work in rectilinear sliding conditions where the effects of these poroelastic flows are minimized. To do so, the load and velocity conditions were chosen so that the Péclet number, which compares the advective (sliding) and poroelastic (drainage) flows within the hydrogel network, is kept lower than 1. The indentation depth $\delta$ is then expected equal to the static one, $\delta_0$, namely poroelastic flows forced by the translation of the sphere on the gel film do not induce any lift force.\\

For both contact conditions, friction experiments were carried out under imposed normal load $F_n$ (between 20 and 200~\si{\milli\newton}) and sliding velocity $V$ (between 0.4 and 450~\si{\micro\meter\per\second}) with the contact fully immersed in deionised water. In addition to monitoring the normal force $F_n$ and measuring the lateral force $F_t$, Reflection Interference Contrast Microscopy (RICM) observations of the immersed contact allowed to continuously monitor the contact radius $a$ and the associated indentation depth $\delta=a^2/2R$.~\cite{delavoipiere_poroelastic_2016, ciapa_transient_2020} Friction experiments involved two steps. In a first step, the contact is allowed to achieve the poroelastic indentation equilibrium (within a few tens of seconds) under a purely static indentation load $F_n$. Then, motion is initiated at imposed velocity $V$ while keeping constant $F_n$.
%
\section{Results}
\subsection{Indentation depth at rest and in steady-state sliding}
\begin{figure}[!ht]
	\includegraphics[height=6cm]{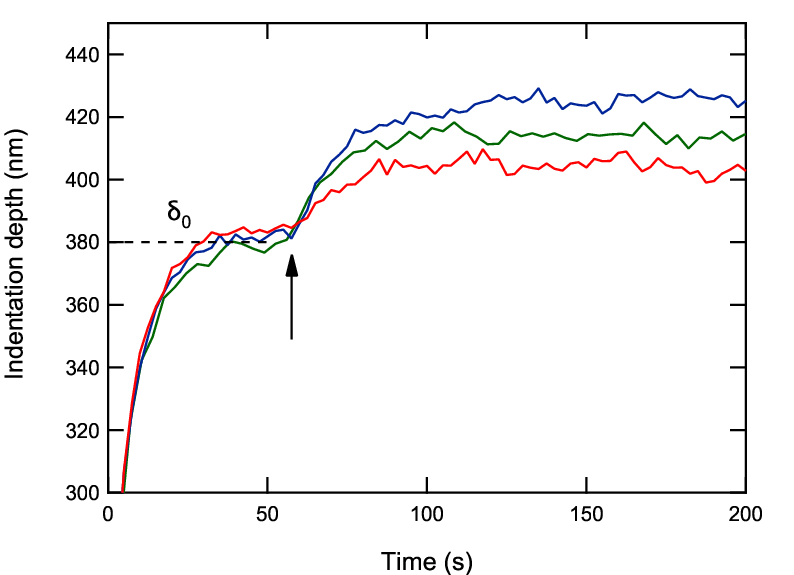}
	\caption{Indentation depth as a function of time $t$ for different functionalized silica spheres (sliding velocity  $V$=240~\si{\micro\meter\per\second}). (blue) propyl, (green) octadecyl, (red) amino. Initial time $t=0$ corresponds to the application of a constant normal force $F_n=100$ mN which leads to a first plateau $\delta_0=380$~nm (dashed line) reached within the poroelastic time $\tau_{\mathrm{poro}}=20$~s.
		The sliding of the silica sphere is initiated at a time $t=60$~s marked by the arrow and gives way to a second indentation step up to a plateau value $\delta > \delta_0$ which depends on the silica chemistry and is reached within $\tau_{\mathrm{poro}}$ as well. Hydrogel film $e_0=2.3~\mu$m, $S_{\mathrm{w}}=1.8$.} 
	\label{fig:d_t_reenfoncement}
\end{figure}

\begin{figure}[!ht]  
	\includegraphics[width=8cm]{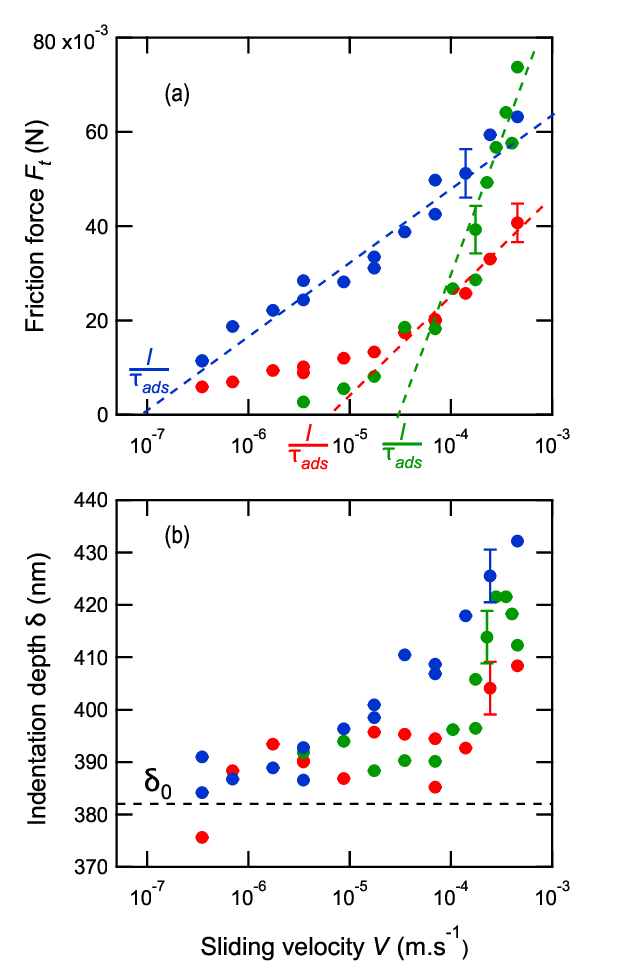}
	\caption{Steady state friction force $F_t$ (a) and indentation depth $\delta$ (b) as a function of sliding velocity $V$ for normal load $F_n=100$~\si{\milli\newton} and different modified silica lenses : (blue) propyl, (green) octadecyl, (red) amino. Lin-log scale. Rotation setup. Dashed lines in (a) are fits to Eq. \ref{eq:sigmat:model:singh} with the intercept to the velocity axis corresponding to $\ell/\tau_{\mathrm{ads}}$. Dashed line in (b): averaged equilibrium indentation depth $\delta_0$ in static condition ($V=0$). Hydrogel film $e_0=2.3~\mu$m, $S_{\mathrm{w}}=1.8$.} 
	\label{fig:ft_d_v}
\end{figure}

At rest, for sliding velocities $V=0$, the indentation depth $\delta_0$ of the sphere into the gel results from a balance between the elastic response of the gel network and the applied constant normal load $F_n$. This balance was detailed elsewhere \cite{delavoipiere_poroelastic_2016} both theoretically and experimentally and we only report the results here. The indentation depth under load at equilibrium is written as :

\begin{equation}
	\delta_0=\left(\frac{e_0F_n}{\pi R \tilde{E}}\right)^{1/2}
	\label{eq:delta_0}
\end{equation}
where $\tilde{E}$ is the oedometric elastic modulus which characterizes the uniaxial compression response of the compressed film.
Experimental indentation data agree with Eq.~\ref{eq:delta_0} with $\tilde{E}=20$~MPa 

and $\tilde{E}=31$~MPa 

respectively for the two hydrogel films.
Indentation equilibrium is achieved within the period of time needed to release the extra pore pressure of the solvent trapped in the gel: this poroelastic time is set by the permeability $\kappa$ of the network for this solvent (viscosity $\eta$) and its elastic modulus $\tilde{E}$ as: $\tau_{\mathrm{poro}}=\frac{R\delta_0}{2\kappa \tilde{E}}$. In this view, the steady-state indentation of the sphere into the hydrogel film depends on the normal load only, for a given experimental system.\\ 
By carefully imaging the contact under rotational sliding conditions, we find instead that the onset of sliding of the sphere relative to the film causes a re-embedding of the sphere into the gel from the equilibrium indentation depth $\delta_0$ which occurs over a time of the order of the poroelastic time $\tau_{\mathrm{poro}}$ (Fig.~\ref{fig:d_t_reenfoncement}). Similar observations are made under rectilinear sliding for low enough velocities $V$ for which the indentation depth is expected to remain equal to the static one, namely poroelastic flows forced by the translation of the sphere on the gel film do not induce any change in the contact load-carrying capacity (cf section 3). Hence, the transient regime observed here in response to sliding fully defers from the lift-induced transients studied in the past.~\cite{ciapa_transient_2020}\\
%
In the following, $\delta$ denotes the steady-state sliding indentation depth. As detailed in Fig.~\ref{fig:ft_d_v}b), the steady sliding indentation depth $\delta$ increases with velocity, i.e. with the value of the steady-state frictional force $F_t$. Noticeably $\delta$ is also dependent on the physical-chemistry of the lens, as marked by the different colors in Figs.~\ref{fig:d_t_reenfoncement} and ~\ref{fig:ft_d_v}b .\\
%
This re-indentation seems to be the manifestation of an additional normal force $\Delta F_n$ emerging with the frictional stress: as we are working with an imposed normal force, we are measuring a variation in indentation rather than a variation in normal force. We calculate $\Delta F_n$ from the difference in indentation between the indentation equilibrium $\delta_0$ and the steady-state frictional regime where indentation is $\delta$. Hence, the indentation depth in steady-state sliding $\delta$ is recast into an increment of normal force $\Delta F_n$. To do so, we use the linear relationship between normal force and indentation depth at equilibrium (Eq.~\ref{eq:delta_0}) :
\begin{equation}
	F_n = \frac{\pi R \tilde{E}}{4 e_0} \delta_0^{2}
	\label{eq:module:tangent}
\end{equation}
so that the additional normal force $\Delta F_n$ is written as:
\begin{equation}
	\Delta F_n=\frac{ \pi R \tilde{E} }{ e_{0}}  \delta^2 - F_n =  F_n\left[\left(\frac{\delta}{\delta_0}\right)^2 - 1\right]
	\label{eq:fnreenf}
\end{equation}

\begin{figure}[!ht]
	\centering
	\includegraphics[height=7cm]{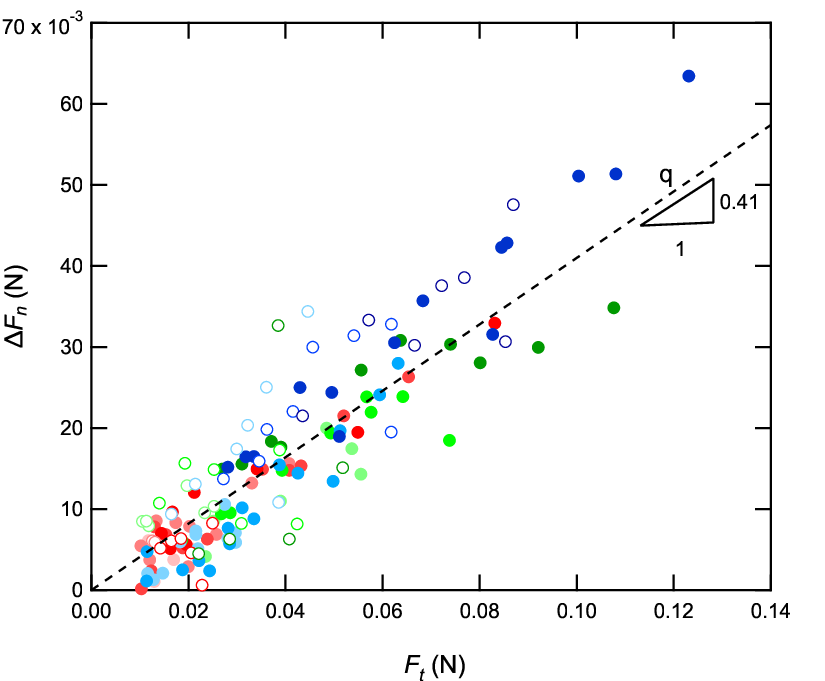}
	\caption{Additional normal force $\Delta F_n$ as a function of the friction force $F_t$ for the different functionalized silica surfaces (blue) propyl, (red) amino and (green) octadecyl, under varied normal load $F_n$ in both rectilinear sliding (empty markers) and rotation sliding (full markers). Light to heavy colors denote applied normal loads $F_n$ ranging from 50 to 200~\si{\milli\newton}. Hydrogel film : swelling ratio $S_{\mathrm{w}}=1.8$; swollen thickness $e_0=2.3~ \mu$m. The dotted line is a linear fit with a slope $q=0.41$.}
	\label{fig:deltafn_ft}
\end{figure}

The resulting $\Delta F_n$  values are plotted in Figure~\ref{fig:deltafn_ft} as a function of the friction force, for all the considered silica chemistries, velocities $V$, normal loads $F_n$, and for the two experimental set-ups, namely rectilinear and rotational sliding : We find that all the data collapse on a single line with a slope measured to be $q=0.41$. We conclude here that a normal stress develops (i) in response to the onset of tangential stresses, (ii) it is compressive and pulls the interface within the hydrogel, (iii) linearly with the friction stress. \\

Let us first discuss this observation in light of the literature. The onset of normal stresses during the shearing of elastic solids in domains of high deformation is a phenomenon that has been known since Poynting's work about a century ago~\cite{poynting1909,poynting1913,zurlo2020} - especially for rubbers - and which was described within the framework of non-linear elasticity by Rivlin.~\cite{rivlin1948} When a neo-Hookean rubber block is sheared between two planes separated by a constant gap, Rivlin's theory predicted that a negative compressive stress, or “dilatancy” is induced. In addition to the fact that this stress has opposed sign to that observed, the theoretical description shows that it is a {\it quadratic} function of the deformation $\gamma$, whereas shear stress varies in $\gamma$. In a later work, Gent and co-workers~\cite{gent2007} showed from finite element calculations that edge effects at the extremities the sheared rubber blocks can induce a change in the sign of the normal stress which becomes tensile. However, this tensile stress remains a {\it quadratic} function of the deformation, \textit{i.e.} a second order effect that only occur at high strains.\\
More recently, Vahabi~\textit{et al.}~\cite{vahabi2018} and Cagny~\textit{et al.}~\cite{decagny2016} have investigated experimentally the development of normal stresses in the case of porous actin networks and poly(acrylamide) gels submitted to a constant shear stress in a constant-gap rheometer. They showed that the normal stress relaxes from an initially compressive or close-to-zero value to a tensile steady-state value. These tensile stresses where observed to develop during a characteristic poroelastic time corresponding to the drainage of the network, which undergoes radial contraction under the action of the orthoradial stress generated by shear. However, the experiments and theoretical model developed by the authors show that these effects are also in $\gamma^2$ and, in the case of poyl(acrylamide) gels, only exist at large strain $\gamma > 1$. As a comparison, the shear strain $\gamma \approx \sigma_t/G$ experienced by the hydrogel films in our experiments do not exceed 0.3. The above reported non linear effects alone cannot therefore account for the proportionality we observe between $\Delta F_n$ and $F_t$, given that $F_t$ varies in $\gamma$.\\
Finally, any possible lift due to the surrounding water advected by the silica sphere rotation would give a stress of opposite sign as what we measure. Indeed, Vialar~\textit{et al.} have shown that shearing of mica surfaces coated by compliant microgel layers can generate a lift force of elastohydrodynamic origin, triggering the entrainment of a fully developed fluid film at large enough speeds, minimizing the contact between the opposite surfaces.~\cite{vialar_compliant_2019}\\

\subsection{Friction stress in the thermally-activated regime}
\begin{figure}[!ht]
	\includegraphics[height=8cm] {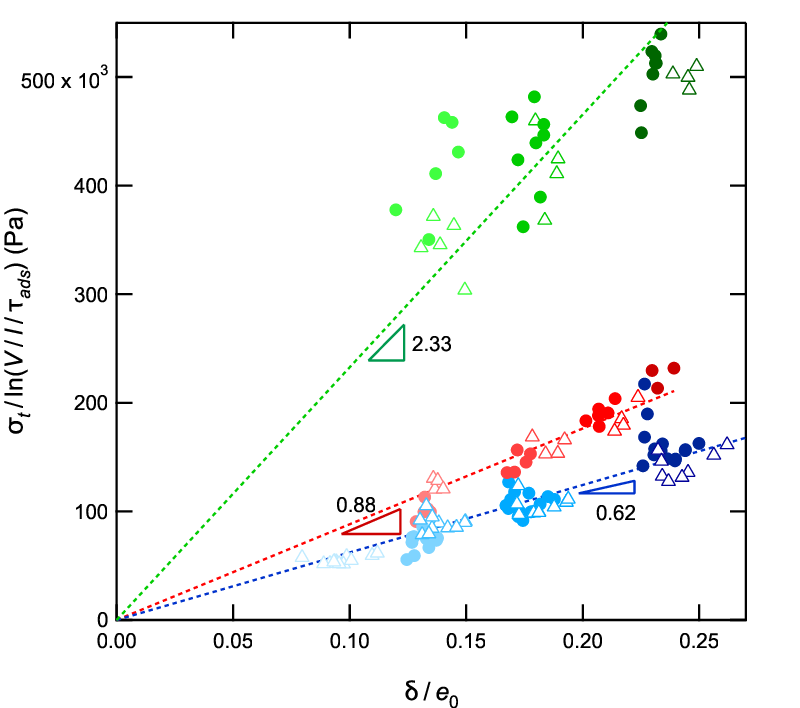}  
	\caption{Steady-state friction stress $\sigma_t$ normalized by the thermally-activated velocity dependent term $\ln(V\tau_{\mathrm{ads}}/\ell)$ as a function of the steady-state indentation depth $\delta$ normalized by the film swollen thickness $e_0$. The ratio $\delta/e_0$ is the normal compressive strain of the film. Light to dark colors : Applied normal load $F_n$ ranging from 50 t0 200~\si{\milli\newton}.   
		Data are collected on two different films differing in thickness and swelling ratio (dots: $e_0=$2.3~\si{\micro\meter}, $S_{\mathrm{w}}=1.8$; triangles: $e_0=$2.8~\si{\micro\meter}, $S_{\mathrm{w}}=2.3$) and various functionalized silica surfaces : all the data collected with the same surface chemistry collapse on a line with a slope denoted $s$ (blue: propyl, $s=0.62$~\si{\mega\pascal}; red: amino, $s=0.88$~\si{\mega\pascal}; green: octadecyl, $s=2.33$~\si{\mega\pascal}).
	}
	\label{fig:sigma_logNorm_indentation}
\end{figure}

We now examine the friction stress versus velocity curves in light of the primary model Eq.~\ref{eq:sigmat:model:singh} where dissipation occurs through molecular adsorption-stretching-desorption of the polymer chains at the sliding interface. Following this prediction, for each tested surface chemistry and for the two sliding geometries, we identify a range of velocity where the stress varies logarithmically with $V$. Examples are shown in Fig.~\ref{fig:ft_d_v}a as dotted lines, where the intercept with the x-axis allows to measure the ratio $\ell/\tau_{\mathrm{ads}}$. As reported in a previous paper~\cite{ciapa_friction_2024}, for each surface chemistry, the data can successfully be described by a single value of the parameter $\ell/\tau_{\mathrm{ads}}$, consistently with the fact that $\tau_{\mathrm{ads}}$ characterizes the chemistry of the bond. On the contrary, the prefactor of the $\ln{V}$ term in Eq.~\ref{eq:sigmat:model:singh}, namely $N_0*\frac{k_BT}{2\lambda}$, is found to depend on the normal load. In order to illustrate this comparison with Eq.~\ref{eq:sigmat:model:singh}, we plot all the friction data in Fig.~\ref{fig:sigma_logNorm_indentation} as a function of the ratio between $\sigma_t$ and $\ln{\frac{V}{\ell/\tau_{\mathrm{ads}}}}$ as a function of the indentation depth $\delta$ normalized by the swollen film thickness $e_0$ for the two different hydrogel films. For each chemistry, all the data collapse onto a single master line which allows to offer a semi-empirical version of the friction law :
\begin{equation}
	\sigma_t=s\frac{\delta}{e_0}\ln{\frac{V}{\ell/\tau_{\mathrm{ads}}}}
\end{equation}
where $s$ has the dimension of a stress and depends on the surface chemistry \cite{ciapa_friction_2024}.

Since $\delta /e_0$ is a measure of the normal strain in the hydrogel film, the linear dependency on $\delta /e_0$ appears as a second evidence of a coupling between the normal and friction stresses. 

\subsection{Normal and tangential stresses couplings in sliding on hydrogel film}

Altogether, our experimental results offer two evidences of a coupling between normal and tangential stresses arising at the onset of sliding and summarized by the three following semi-empirical equations : (A) Linear relationship between stresses Eq.~\ref{eq:reenfoncement_exp},  (B) Friction stress dependence on indentation and velocity Eq.~\ref{eq:sigmafitlogV_delta}, and (C) Normal equilibrium Eq.~\ref{eq:fnreenf} which is Eq.\ref{eq:reenfoncement_exp} recast into $\delta$ using $\pi a^2=2\pi R \delta$ as the contact area:

\begin{equation}
	\mathrm{(A)}\indent \frac{\Delta F_n}{F_t}=\frac{\Delta\sigma_n}{\sigma_t}=q\sim0.4
	\label{eq:reenfoncement_exp}
\end{equation}
\begin{equation}
	\mathrm{(B)} \indent
	\sigma_t=s\frac{\delta}{e_0}\ln{\frac{V}{\ell/\tau_{\mathrm{ads}}}} 
	\label{eq:sigmafitlogV_delta}
\end{equation}
\begin{equation}
	\mathrm{(C)} \indent F_n=\frac{\pi R \tilde{E}}{e_{0}} \delta^2-\Delta \sigma_n 2\pi R \delta
	\label{eq:fnreenf}
\end{equation}

In the following, we derive a revised version of the thermally-activated model for friction by building upon an idea offered by Leonov to account for the non-zero thickness of the interface from which polymer chains adsorb. We then compare the prediction to the experimental results (A), (B) and (C).

\section{Discussion}

\subsection{Molecular friction model for finite non-zero interfacial thicknesses}

\begin{figure}[!ht]
	\centering
	\includegraphics[height=4cm]{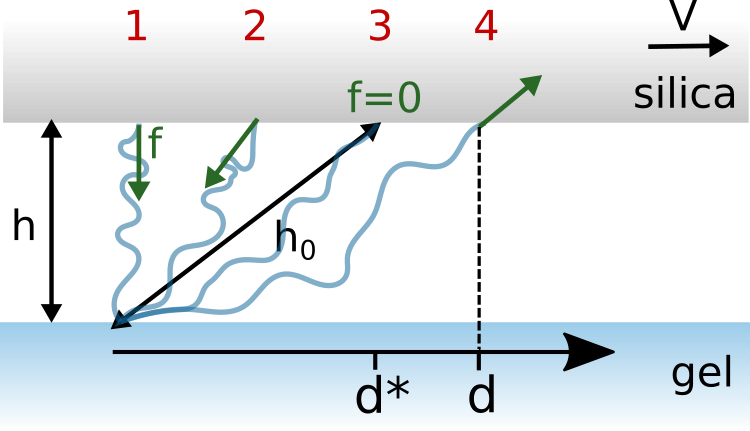}
	\caption{Schematic representation of a polymer chain adsorbed at the sliding surface of silica having a velocity $V$. $d$ is the distance between the adsorption point and the crosslink point of the chain to the gel network. $f$ is the force acting on a chain by the sliding silica interface (1) $d=0$, $f<0$ : under normal load $F_n$, the polymer chain is compressed to a length $h$ smaller than the equilibrium length $h_0$ of the polymer chain. The force is repulsive. 2) $0<d<d^*$; $f<0$. 3) $d=d^*$, $f=0$ and the chain is at its equilibrium length $h_0$. 4) $d>d*$, $f>0$ : the force $f$ stretches the chain out of the plane of the interface and is attractive. We assume that under slding, readsorption occurs when $f=0$, that is in case 3, for which the chain length is $h_0$. 
	}
	\label{fig:ressorts_inclines_a}
\end{figure}

\begin{figure}[!ht]
	\vspace{10pt}
	\centering
	\includegraphics[height=4cm]{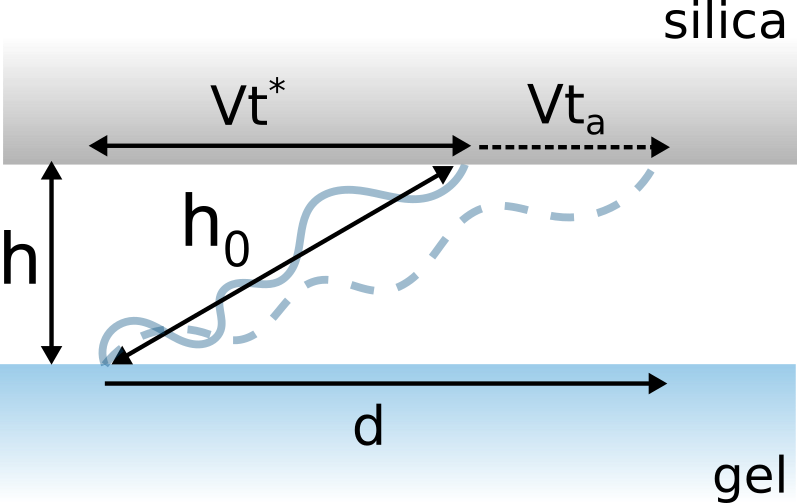}
	\caption{Schematic representation of a polymer chain adsorbing at the sliding surface of silica having velocity $V$ (line) and being further stretched (dashed line). Adsorption occurs when $f=0$ so that the chain is necessarily inclined, with a non-zero distance $d^*=Vt^*$ between the adsorption point and the chain anchor in the gel. As further sliding occurs, the adsorption time $t_a$ increases and the distance $d$ increases accordingly: $d=d^*+Vt_a$. The tilt of the chain with respect to the normal direction to the interface is characterized by $x^*=Vt^*/h$. 
	}
	\label{fig:ressorts_inclines}
\end{figure}
Let's return to the primary adsorption/desorption model described by Eq.~\ref{eq:sigmat:model:singh} which assumes that the force $f$ exerted by the adsorbed surface polymer chains is parallel to the rigid interface, or equivalently, that the interface has a vanishing thickness $h \rightarrow 0$. However, if the point of attachment of the surface molecules to the network is at a non zero distance $h$ from the sliding surface, as shown in Fig. \ref{fig:Schallamach_leonov}, then the force $f$ not only has a tangent component at the interface, but also a normal component. Leonov \cite{chernyak_theory_1986,leonov_dependence_1990} has proposed an elastomer friction model adapted from Schallamach's model and taking into account this interface geometry where the surface layer has a fixed thickness $h$. Leonov's model provides an expression not only for the frictional stress $\sigma_t$ but also for the normal stress $\sigma_n$ resulting from the normal component of the force $f$ exerted by each of the adsorbed chains. At a fixed distance $h$, Leonov shows that friction induces a decrease in normal stress, which would explain a re-embedding phenomenon in our imposed normal force geometry. The variation in normal stress is substantial here, since Leonov calculates a decrease in $\sigma_n$ of the order of ten percent as the velocity increases by a factor of ten, in the logarithmic velocity friction regime. However, although Leonov's model highlights a coupling between normal and tangential stresses, the calculated ratio between these two stresses is velocity-dependent, in contrast to our experimental observations, which show a constant ratio $q$ of around 0.5. Furthermore, Leonov's approach does not consider the pre-compression of chains due to the application of a normal force at zero sliding speed.\\
To characterize the effects of the applied normal force we have measured, the latter must be considered. In the following, we propose a calculation inspired by Leonov's description and Singh's model described above, and taking into account an imposed normal force. We shall see that this model enables us to discuss a possible mechanism giving rise to a coupling between normal stress and frictional stress.\\

\subsubsection*{Adsorption/stretching/desorption of inclined polymer chains}

If a chain adsorbed on the sliding surface is prestressed in compression, the chain has a length less than its equilibrium length, which we note $h_0$, up to a certain displacement of the adsorption point: in this case the force $f$ is repulsive and not attractive as shown in Fig. \ref{fig:ressorts_inclines_a}, which would give rise to a negative friction force. We then propose a model in which the surface chains adsorb in the inclined state, with a length at least equal to the equilibrium length. Note $h_0$, the characteristic size of surface molecules at rest. The force exerted after a displacement $d$ of the end of the molecule is written in the linear approximation :

\begin{equation}
	f(d)=M\left[\sqrt{h^2+d^2}-h_0\right]
\end{equation}
$h$ represents the thickness of the surface layer under friction. This force cancels out for a displacement $d^*$ :
\begin{equation}
	d^*=\sqrt{h_0^2-h^2}
\end{equation} 
For $d<d^*$, the interfacial force is globally repulsive, whereas it is attractive in the opposite case. We therefore consider that the chains adsorb in an inclined state characterized by $d=d^*$. We note $d^*=Vt^*$ to describe that from the moment of attachment ($t=0$), the reaction force is written as:
\begin{equation}
	f(t_a)=M\left[\sqrt{h^2+V^2(t_a+t^*)^2}-h_0\right] \label{eq:ressort}
\end{equation} 
with $Vt^*=\sqrt{h_0^2-h^2}$ and $t_a$ the age of the bond. In this description, the chains re-adsorb with zero interfacial force, but in an inclined position. The situation is illustrated in Fig. \ref{fig:ressorts_inclines}. In particular, we maintain the assumption that the transition from a free chain to a bond is not activated, takes place at zero force, and the free state is still characterized by a lifetime $\tau$.
The steady-state surface layer thickness, $h$, is not specified {\it a priori} in the model, but will be obtained from a normal equilibrium condition that combines the response of all the molecules and the normal force applied. Re-embedding is considered to be caused by the normal contribution $f_n$ of the bonding force. It is assumed here that the majority of chains on the surface withstand the normal force $F_n$, while only a small number of chains bond and give rise to friction and re-embedding.

By projection, the normal and tangential components of the binding force are
\begin{align}
	f_t(t_a)&=\frac{V(t_a+t^*)}{\sqrt{h^2+V^2(t_a+t^*)^2}}f(t_a)\\
	f_n(t_a)&=\frac{h}{\sqrt{h^2+V^2(t_a+t^*)^2}}f(t_a)
\end{align}

The survival probability $P_{V,h}(t_a)$ for a bond formed at a time $t=0$ to exist after a time $t_a$ is modified into:
\begin{equation}
	P_{V,h}\left( t_a \right)=\exp\left( -\frac{1}{\tau_{\mathrm{ads}}}\int_0^{t_a}e^{\lambda f\left( \xi \right)/kT}d\xi \right)
	\label{eq:gta}
\end{equation} 
so that the distribution function $n(t_a)$ of the number of bonds aged $t_a$ is:
\begin{equation}
	n\left( t_a \right)={N_0}\frac{\frac{1}{\tau} P_{V,h}\left( t_a \right)}{1+\frac{1}{\tau}\int_0^{\infty}P_{V,h}(t)dt}
\end{equation} 
Using the same description as in reference\cite{ciapa_friction_2024}, the stresses $\sigma_i$ can be written as a function of  $P_{V,h}$ as $\sigma_i=\int^{\infty}_0 n(t_a)f_i(t_a) dt_a$:
\begin{align}
	\sigma_t&={N_0}\frac{\frac{1}{\tau}\int_0^\infty  P_{V,h}\left( t_a \right)f_t\left( t_a \right)dt_a}{1+\frac{1}{\tau}\int_0^\infty  P_{V,h}\left( t_a \right)dt_a}\\
	\Delta\sigma_n&={N_0}\frac{\frac{1}{\tau}\int_0^\infty  P_{V,h}\left( t_a \right)f_n\left( t_a \right)dt_a}{1+\frac{1}{\tau}\int_0^\infty  P_{V,h}\left( t_a \right)dt_a}
	\label{eq:model:sigmatn:0}
\end{align}
We are aiming at modelling the three experimental observations (A, B, C) described by the equations~\ref{eq:fnreenf}, \ref{eq:reenfoncement_exp} and \ref{eq:sigmafitlogV_delta}.

\begin{equation}
	P_{V,h}\left( t_a \right)=\exp\left( -\frac{1}{\tau_{\mathrm{ads}}}\int_0^{t_a}e^{\left[\sqrt{h^2+V^2(t+t^*)^2}-h_0\right]/\ell}\,dt \right)
\end{equation} 
We define:
\begin{align}
	Q&=\frac{h}{\ell}\\
	u'&=\frac{h}{V\tau_{\mathrm{ads}}}
\end{align}
so that :
\begin{equation}
	P_{V,h}\left( t_a \right)=\exp\left( -u'\int_0^{Vt_a/h}e^{Q\left(\sqrt{1+(y+x^*)^2}-\sqrt{1+x^{*2}}\right)}\,dy \right)
\end{equation} 
with
\begin{equation}
	x^*=Vt^*/h=\sqrt{\left( \frac{h_0}{h} \right)^2-1}
	\label{eq:xstrarh}
\end{equation}
where $x^*$ characterizes the tilt angle $\tan^{-1}(x^*)$ of the newly attached chain with respect to the normal direction at the interface (Fig.~\ref{fig:ressorts_inclines}).
The stresses given by Eqs.~\ref{eq:model:sigmatn:0} can then be further derived by substituting the integration variable $t_a$ with $x=Vt_a/h$ and denoting $\tilde{\phi}(x)=\phi(t_a)$.
\begin{align}
	\frac{h_0}{h}&=\sqrt{1+x^{*2}}\\
	\tilde f_t(x)&=Mh\frac{x+x^*}{\sqrt{1+(x+x^*)^2}}\left[\sqrt{1+(x+x^*)^2}-\sqrt{1+x^{*2}}\right]\\
	\tilde f_n(x)&=Mh\frac{1}{\sqrt{1+(x+x^*)^2}}\left[\sqrt{1+(x+x^*)^2}-\sqrt{1+x^{*2}}\right]\\
	\tilde P_{V,h}(x)&=\exp\left( -u'\int_0^{x}e^{Q\left(\sqrt{1+(y+x^*)^2}-\sqrt{1+x^{*2}}\right)}\,dy \right)
	\label{eq:gtilde}
\end{align}
and
\begin{equation}
	\sigma_t={N_0}\frac{h}{V\tau}\frac{\int_0^\infty \tilde P_{V,h}\left( x \right)\tilde f_t\left( x \right)\,dx}{1+\frac{h}{V\tau}\int_0^\infty \tilde P_{V,h}\left( x \right)\,dx}
	\label{eq:contrainte:t}
\end{equation}

\begin{equation}
	\Delta\sigma_n={N_0}\frac{h}{V\tau}\frac{\int_0^\infty \tilde P_{V,h}\left( x \right)\tilde f_n\left( x \right)\,dx}{1+\frac{h}{V\tau}\int_0^\infty \tilde P_{V,h}\left( x \right)\,dx}
	\label{eq:contrainte:n}
\end{equation}

Note that the ratio $Q=h/\ell$ is of order 1 since it compares two molecular lengths, both related to the network mesh size ($h\sim \ell \sim 10$~nm). In the logarithmic regime for friction, which is the object of the present study, velocity ranges between $\ell/\tau_{\mathrm{ads}}\ll V \ll \ell/\tau$ so that $\frac{h}{V\tau_{\mathrm{ads}}}\ll Q \ll \frac{h}{V\tau} $, so that $u' \ll 1$ and $h \gg V \tau$.

\subsubsection*{Tilt angle of the chains $\tan^{-1}(x^*)$}
The length ratio $x^*$ is set by Eq.~\ref{eq:xstrarh} and depends on the stretch $h_0/h$. The latter is estimated from the compressive strain $1-h/h_0$ of the interfacial layer by assuming it increases linearly with the compressive strain of the film $\delta/e_0$. This allows to introduces a degree of freedom accounting for a possible difference in compressibility between the interfacial dangling chains and the bulk cross-linked ones. Hence, we define the ratio $k$ between the elastic moduli of the superficial layer and the hydrogel and $k$ is of order and smaller than 1. It follows that: 
\begin{equation}
	\frac{h_0-h}{h_0}=\frac{1}{k}\frac{\delta}{e_0}
	\label{eq:xh0h}
\end{equation} 
For typical values of $\delta/e_0\sim0.2$ taken from Fig.~\ref{fig:sigma_logNorm_indentation}, $k\sim1$ to 0.75, we find $x^*=0.7$ to 1 which means that the polymer chains are significantly tilted : the tilt angle is $\tan^{-1}{x^*}\simeq 25^\circ$ to 45$^\circ$. In the following, we will assume $x^*\sim 1$ which holds except in the low force range.

\subsubsection*{Derivation of the normal and friction stresses}

Recalling that $u'\ll1$, $x^*\sim 1$ and $h$ ranges between 1~nm to 10~nm, we will now estimate the stresses. First, in the velocity range $\ell/\tau_{\mathrm{ads}}\ll V \ll \ell/\tau$, Eq.~\ref{eq:gtilde} can be simplified by following the path of the previous derivation \cite{ciapa_friction_2024} and by noting that $\tilde{P}_{V,h}(x)$ is a step-down function with a cut-off value $X$ for $x$ defined by:
\begin{equation}
	\int_0^Xe^{Q\left(\sqrt{1+(y+x^*)^2}-\sqrt{1+x^{*2}}\right)}\,dy=\frac{\ln2}{u'}
\end{equation} 
With $u'\ll 1$ and $x^*\sim 1$, we can show that $X$ is well approximated by \footnote{With $u'\ll 1$ and $x^*\sim 1$, we find that $X+x^*\gtrsim1$ so that the integral in Eq.~\ref{eq:gtilde} is mostly set by the $(y+x^*)$ values larger than 1 so that $X$ approximately verifies: $e^{Q(x^*-\sqrt{1+x^{*2}})}\int_0^Xe^{Qy}\,dy=\frac{\ln2}{u'}
	$}:
\begin{equation}
	X\simeq \sqrt{1+x^{*2}}-x^*+\frac1Q\ln\frac{Q}{u'}\label{eq:X}
\end{equation} 
The cut-off value $X$ is next used to estimate the integrals in Eqs.~\ref{eq:contrainte:t}, \ref{eq:contrainte:n} by replacing the upper bound of integration $\infty$ by $X$.      
	In the following, we use again the assumption that $h \gg V\tau$ : in the thermal regime, the chain are barely strained by shear over the time $\tau$. We assume in the following that $X\ll 1$ :
	\begin{align}
		\sigma_t&\simeq \frac12 {N_0}Mh\frac{x^{*2}}{1+x^{*2}}X\label{eq:deltasigmatX}\\
		\Delta \sigma_n&\simeq \frac12 {N_0}Mh\frac{x^{*}}{1+x^{*2}}X
	\end{align}
	We directly recover the experimental result (A) described by the empirical Eq.~\ref{eq:reenfoncement_exp} with :
	\begin{equation}
		\frac{\Delta\sigma_n}{\sigma_t}=\frac{1}{x^*}
	\end{equation} 
	where we identify $q$ with $1/x^*$. Experimental data further show that $q$ depends neither on velocity nor on normal load, within experimental accuracy, so that the agreement between the data and the model requires that $x^*$ should be independent on $V$ and $F_n$. An exact resolution of the problem would allow to further validate this point. The data from Fig.~\ref{fig:deltafn_ft} provides a measure of $q=0.41$ so that $x_{exp}^*=2.5$.\\
	
	Let us now examine the experimental property (B). To do so, we further derive the friction stress so as to specify the velocity $V$ logarithmic dependence as well as the $\delta/e_0$ linear dependence as measured by Eq.~\ref{eq:sigmafitlogV_delta}. First, note that $\delta/e_0$ is of the order of $(h_0-h)/h_0$ through Eq.~\ref{eq:xh0h}. We then approximate $X$ given by Eq.~\ref{eq:X} by considering the limit where $x^*$ is large enough so that $\sqrt{1+x^{*2}}-x^*$ can be neglected compared to the second term: experimentally, $x_{exp}^*=2.5$ so that $\sqrt{1+x^{*2}}-x^*=0.2$ while $Q/u'$ varies between 10 and $10^4$ so that the natural logarithm term is always larger than 3. Eq.~\ref{eq:X} is approximated as:
	
	\begin{equation}
		X\simeq \frac 1Q\ln\frac{Q}{u'} = \frac{\ell}{h}\ln\frac{V}{\ell/\tau_{\mathrm{ads}}}
	\end{equation} 
	so that
	\begin{align}
		\sigma_t&\simeq \frac{{N_0} k_BT}{2\lambda}\frac{x^{*2}}{1+x^{*2}}\ln\frac{V}{\ell/\tau_{\mathrm{ads}}}\label{eq:deltasigmat}\\
		\Delta \sigma_n&\simeq \frac{{N_0} k_BT}{2\lambda}\frac{x^*}{1+x^{*2}}\ln\frac{V}{\ell/\tau_{\mathrm{ads}}} \label{eq:deltasigman}
	\end{align}
	The geometric prefactor in $\sigma_t$ is
	$ \frac{x^{*2}}{1+x^{*2}}=\frac{h_0^2-h^2}{h^2}\frac{h^2}{h_0^2}=\frac{(h_0-h)(h_0+h)}{h_0^2} \simeq \frac{\delta}{ke_0}\frac{h_0+h}{h_0}\simeq\frac{2\delta}{ke_0}$, assuming that $h_0-h \ll h_0$. The friction stress is finally written:
	\begin{equation}
		\sigma_t=\frac{{N_0} k_BT}{k\lambda}\frac{\delta}{e_0}\ln\frac{V}{\ell/\tau_{\mathrm{ads}}}
	\end{equation} 
	which corresponds exactly to the semi-experimental property (B) given by Eq.~\ref{eq:sigmafitlogV_delta} where we identify $s$ with $\frac{{N_0} k_BT}{k\lambda}$. The number density of bonds ${N_0}$ depends on the surface chemistry. It is of the order of 1 to 1/10 per nm$^2$ and its variation with the surface chemistry of the silica lens was commented in our earlier work.~\cite{ciapa_friction_2024} The modulus ratio $k$ depends on the hydrogel film and will be commented later.\\
	
	We finally turn to the empirical property (C) that links the additional normal stress $\Delta \sigma_n$ to the applied normal load $F_n$. This link is accounted for by Eq.~\ref{eq:fnreenf} which is a second order equation in $\delta$ if we remember that measurements are done under set load conditions. We offer then to solve this equation and compare the resulting $\delta$-values with experimental data from Fig.~\ref{fig:ft_d_v}b while using the predicted values of $\Delta\sigma_n$ from the model Eq.~\ref{eq:deltasigman}. The second order equation has two solutions, one being physically impossible ($10^{-18}$ m) and the other being written as:
	\begin{equation}
		\delta=\frac{e_0\Delta \sigma_n}{\tilde{E}} \left( 1+ \frac{1}{2}\sqrt{1+\frac{ F_n \tilde{E}}{\pi R \Delta \sigma_n^2 e_0}} \right)
		\label{eq:solution:delta}
	\end{equation}
	
		From the friction data with the propyl-silanated lens, we measure in Fig.~\ref{fig:sigma_logNorm_indentation}: $s\equiv \frac{{N_0} k_BT}{k\lambda} =0.62$~\si{\mega\pascal}, and in Fig.~\ref{fig:ft_d_v} : $\ell/\tau_{\mathrm{ads}}\sim5.10^{-8}$ m.s$^{-1}$. With $x_{\mathrm{exp}}^*=2.5$ and $k=1$, 
		the model predicts from Eq.~\ref{eq:deltasigman} that $\Delta \sigma_n $ varies between 0.4 and 1.8~\si{\mega\pascal} for velocities ranging between 4.10$^{-7}$ and 4.10$^{-4}$~m.s$^{-1}$. From these estimates, by using Eq.~\ref{eq:solution:delta} and experimental values : $F_n=100$~mN, $R=23$ mm, $e_{0}=2.3 ~ \mu$m, $S_{\mathrm{w}}$=1.8, 
		$\tilde{E}\sim 20.10^6 $ Pa,
		we obtain $\delta$ values predicted between 300 and 480~nm, in fair agreement with experimental data reported in Fig.~\ref{fig:ft_d_v}b. This allows to validate the agreement between empirical property (C) and the present model. \\
		
		\subsection{Comments on the interfacial thickness and softness}
		
		We finally comment on the hypotheses underlying the derivation of the model. We assumed $X \ll 1$. From Eq.~\ref{eq:deltasigmatX}, a value of $hX$ can be obtained from the measured typical value of $\sigma_t\sim 1.10^6$~Pa : we find $hX=20$~nm. Hence, $X$-values smaller than 1 suggest the interface thickness $h$ is larger than 20~nm. Using $x^*_{\mathrm{exp}}=2.5$, an equilibrium length $h_0=50$~nm is found. Such a large value, when compared to the mesh size, suggests that $h$ and $h_0$ are effective lengths which account for both the softness of the gel film which supports the interfacial layer, and the possibly distributed lengths of these superficial chains. An estimate of the superficial layer relative softness $k$ can be obtained from Eq.~\ref{eq:xh0h} using $h_0=50$~nm, $h=20$~nm and measured values of $\delta/e_0\sim 0.15$ : we obtain an elastic moduli ratio $k\sim 0.25$.
				
		It is then possible to convert the elastic modulus of the interfacial layer obtained from $k$ into a polymeric length of the dangling chain and compare it to $h$. To do so, a classical Flory-Rehner model of 1D-swelling is first used\cite{augustine_swelling_2023} to convert the bulk elastic modulus of the swollen network $\tilde{E}$ into a shear modulus of the drained networks $G_0$. For the bulk film, we measure $\tilde{E}^{\mathrm{bulk}}=20$~MPa so that $G_0^{\mathrm{bulk}}=1.9$~MPa which corresponds to a number of monomers between crosslinks of $\nu_{\mathrm{c}}$=13 \footnote{Through $G_0=\rho RT/M_{\mathrm{c}}$ where $\rho=962$~kg/m$^3$ is the density, $M_{\mathrm{c}}=\nu_{\mathrm{c}}M_{\mathrm{k}}$ is the molar mass between entanglements and $M_{\mathrm{k}}=96$~g/mol is the monomer molar mass.}. From $k=0.25$, the elastic modulus of the superficial layer is then $\tilde{E}^{\mathrm{s}}=k \tilde{E}^{\mathrm{bulk}}= 5$~MPa: it corresponds to a swelling ratio of 3 and a drained network shear modulus of $G_0^{\mathrm{s}}=0.4$~MPa so that the equivalent number of monomers is $\nu_{\mathrm{c}}^{\mathrm{s}}=65$. The corresponding Flory diameter of dangling chains is $2b{\nu_c^{\mathrm{s}}}^{0.58}=25$~nm which compares well with the $h$ and $h_0$-values obtained from the friction data. While it appears challenging to further characterize the equilibrium length $h_0$ of the surface chains, one could further test the model by grafting controlled-length chains at the film interface and observe the effect of their size on the normal and frictional stresses.   
		
		Altogether, the assumptions underlying the derivation of the present model were fairly checked experimentally, either by comparison with data from the friction experiments, or by independent characterizations of the gel interface. The analytical derivation we offer allows to provide ready-to-use analytical equations. Nevertheless, a numerical resolution of Eq.~\ref{eq:gtilde}, \ref{eq:contrainte:t},\ref{eq:contrainte:n} could possibly allow to extend the validity of the results out of the limits required by the analytical derivation, in particular the $X<<1$ hypothesis. Hence, the present model offers a mechanism that allows to account for both the additional normal stress arising upon sliding and the friction stress dependence on both normal strain and velocity. By comparison with our experimental data on well-controlled systems, this mechanism is clearly identified to be the adsorption-stretching-desorption of inclined polymer chains from a non-zero superficial layer of the gel network in the thermally activated regime.\\
				
		\section{Conclusions}
		
		The frictional and normal response of a sliding hydrogel-solid interface was carefully measured using a newly developed experimental set-up having the following characteristics. (i) A solid silica sphere is rotated at the surface of a thin hydrogel film (a micrometer-thick, grafted and crosslinked polymeric network of PDMA swollen with water) so that the sliding velocity $V$ can be considered as uniform within the contact. (ii) Poroelastic flows within the gel and entrapment of a lubrication film at the interface are negligible. (iii) The set-up operates at controlled velocity and set normal load, and provides a fine measurement of the frictional force, and of the normal strain, and a direct visualization of the contact with enhanced optical sensitivity. \\
		By working in the low velocity thermal regime \cite{singh_model_2021,ciapa_friction_2024}, the results evidence (i) a dependence of the frictional stress with both the velocity, the normal load, and the physico-chemistry of the interface and (ii) for the first time, a coupling between the frictional and normal response of the interface. This coupling is not due to elasto-hydrodynamics. It results in (i) a friction stress being proportional to the normal strain of the gel and (ii) a normal tensile stress in response to sliding. Based on a newly derived thermally-activated model built upon Schallamach \cite{schallamach_theory_1963} and Leonov \cite{leonov_dependence_1990} models, we ascribe the coupling to the molecular adsorption of polymer chains at the interface between the gel and the solid by accounting for the non-zero thickness of the superficial layer which results in an out-of-plane component of the force exerted between the adsorbed polymer chains and the solid probe. This model successfully allows to semi-quantitatively recover the experimental observations (i) and (ii). \\
		The present work brings experimental evidence of the development of an adhesive-like force at the sliding interface of a gel. In friction experiments conducted at set indentation rather than set load, we anticipate the normal-tangential coupling should result in a decrease in the normal load needed to maintain the gap, between the static and sliding situations. In friction experiments conducted with a gradient of interfacial velocity, for instance using rheometers with rotating plates, the velocity-dependence of both components of the interfacial stress results in gradients of frictional and tensile stresses across the sliding contact. The latter could result in complex couplings between the response of the interface and that of the bulk that would be relevant to account for in a quantitative interpretation of the force-velocity curves that were measured in such experiments. Hence, the tensile stresses induced by molecular adsorption at frictional interfaces should have implications in any friction experiments performed on polymeric gels in the low velocity regime where the population of adsorbed versus desorbed chains is ruled by thermal energy. \\
		Finally, the present system appears as a very rich situation from which friction can be quantitatively related to molecular parameters, at variance with phenomenological descritpions available in the literature. In addition, several physico-chemical parameters could be tested, such as the chemistry of some monomers or the molecular length of the polymer chains at the sliding interface that could be varied by grafting brushes.\\
		
		\section*{Acknowledgements}
		The authors thank Francois Lequeux and Catalin Picu for fruitful discussions. They also would like to thank Joshua McGraw for his insights on swelling instabilities of the gel films.
		
		
		
		\balance
		

\bibliographystyle{rsc} 
\bibliography{biblio.bib} 

\end{document}